\documentclass[11pt,a4paper]{article}
\usepackage{amsmath,amsthm,amsfonts,comment}

\DeclareMathOperator{\Hom}{Hom}
\DeclareMathOperator{\Com}{Com}

\DeclareMathOperator{\1}{id}
\newcommand{\NN}{\mathbb{N}}
\newcommand{\RR}{\mathbb{R}}

\newcommand{\EEnd}{\mathcal End}
\newcommand{\EE}{\mathcal E}
\newcommand{\bul}{\bullet}
\renewcommand{\=}{:=}
\renewcommand{\t}{\otimes}
\renewcommand{\:}{\colon}
\newcommand{\al}{\alpha}
\newcommand{\be}{\beta}
\newcommand{\m}{\overset{\circ}{\mu}}
\newtheorem{thm}{Theorem}[section]
 \newtheorem{lemma}[thm]{Lemma}
 \newtheorem{cor}[thm]{Corollary}
\theoremstyle{definition}
 \newtheorem{defn}[thm]{Definition}
\theoremstyle{definition}
 \newtheorem{exam}[thm]{Example}
\theoremstyle{definition}
 \newtheorem{rem}[thm]{Remark}
\numberwithin{equation}{section}
\begin{document}
\title{\LARGE\bf  3D binary anti-commutative operadic Lax\\
representations for harmonic oscillator}
\author{\Large Eugen Paal and J\"{u}ri Virkepu}
\date{}\maketitle \thispagestyle{empty}

\begin{abstract}
It is explained how the time evolution of the operadic variables may
be introduced by using the operadic Lax equation. The operadic Lax
representations for the  harmonic oscillator are constructed in
3-dimensional binary anti-commutative algebras. As an example, an
operadic Lax representation for the harmonic oscillator in the Lie
algebra $\mathfrak{sl}(2)$ is constructed.

\end{abstract}

\section{Introduction}

In Hamiltonian formalism, a mechanical system is described by the
canonical variables $q^i,p_i$ and their time evolution is prescribed
by the Hamiltonian equations
\begin{equation}
\label{ham} \dfrac{dq^i}{dt}=\dfrac{\partial H}{\partial p_i}, \quad
\dfrac{dp_i}{dt}=-\dfrac{\partial H}{\partial q^i}
\end{equation}
By a Lax representation  \cite{Lax68,BBT03} of a mechanical system
one means such a pair $(L,M)$ of matrices (linear operators) $L,M$
that the above Hamiltonian system may be represented as the Lax
equation
\begin{equation}
\label{lax} 
\dfrac{dL}{dt} = ML-LM
\end{equation}
Thus, from the algebraic point of view, mechanical systems can be
described by linear operators, i.e by  linear maps $V\to V$  of a
vector space $V$. As a generalization of this one can pose the
following question \cite{Paal07}: how to describe the time evolution
of the linear operations (multiplications) $V^{\t n}\to V$?

The algebraic operations (multiplications) can be seen as an example
of the \emph{operadic} variables \cite{Ger}. If an operadic system depends on time one can speak
about \emph{operadic dynamics} \cite{Paal07}. The latter may be
introduced by simple and natural analogy with the Hamiltonian
dynamics. In particular, the time evolution of the operadic
variables may be given by the operadic Lax equation. In
\cite{PV07,PV08-1}, a 2-dimensional binary operadic Lax representation
for the harmonic oscillator was constructed. In the present paper we
construct the operadic Lax representations for the harmonic
oscillator in 3-dimensional binary anti-commutative algebras. As an
example, an operadic Lax representation for the harmonic oscillator
in the Lie algebra $\mathfrak{sl}(2)$ is constructed.

\section{Endomorphism operad and Gerstenhaber brackets}

Let $K$ be a unital associative commutative ring, $V$ be a unital $K$-module, and 
$\EE_V^n\= {\EEnd}_V^n\= \Hom(V^{\t n},V)$
($n\in\NN$). For an \emph{operation} $f\in\EE^n_V$, we refer to $n$
as the \emph{degree} of $f$ and often write (when it does not cause
confusion) $f$ instead of $\deg f$. For example, $(-1)^f\= (-1)^n$,
$\EE^f_V\=\EE^n_V$ and $\circ_f\= \circ_n$. Also, it is convenient to use
the \emph{reduced} degree $|f|\= n-1$. Throughout this paper, we
assume that $\t\= \t_K$.

\begin{defn}[endomorphism operad \cite{Ger}]
\label{HG} 
For $f\t g\in\EE_V^f\t\EE_V^g$ define the \emph{partial compositions}
\[
f\circ_i g\= (-1)^{i|g|}f\circ(\1_V^{\t i}\t g\t\1_V^{\t(|f|-i)}) \quad \in\EE^{f+|g|}_V,
         \quad 0\leq i\leq |f|
\]
The sequence $\EE_V\= \{\EE_V^n\}_{n\in\NN}$, equipped with the partial compositions $\circ_i$, is called the \emph{endomorphism operad} of $V$.
\end{defn}

\begin{defn}[total composition \cite{Ger}]
The \emph{total composition} $\bul\:\EE^f_V\t\EE^g_V\to\EE^{f+|g|}_V$ is
defined by
\[
f\bul g\= \sum_{i=0}^{|f|}f\circ_i g\quad \in \EE_V^{f+|g|}, \quad |\bul|=0
\]
The pair $\Com\EE_V\= \{\EE_V,\bul\}$ is called the \emph{composition algebra} of $\EE_V$.
\end{defn}

\begin{defn}[Gerstenhaber brackets \cite{Ger}]
The  \emph{Gerstenhaber brackets} $[\cdot,\cdot]$ are defined in
$\Com\EE_V$ as a graded commutator by
\[
[f,g]\= f\bul g-(-1)^{|f||g|}g\bul f=-(-1)^{|f||g|}[g,f],\quad |[\cdot,\cdot]|=0
\]
\end{defn}

The \emph{commutator algebra} of $\Com \EE_V$ is denoted as
$\Com^{-}\!\EE_V\= \{\EE_V,[\cdot,\cdot]\}$. One can prove (e.g \cite{Ger}) that $\Com^-\!\EE_V$
is a \emph{graded Lie algebra}. The Jacobi identity reads
\[
(-1)^{|f||h|}[[f,g],h]+(-1)^{|g||f|}[[g,h],f]+(-1)^{|h||g|}[[h,f],g]=0
\]

\section{Operadic Lax equation and harmonic oscillator}

Assume that $K\= \RR$ or $K\= \mathbb{C}$ and operations are
differentiable. Dynamics in operadic systems (operadic dynamics)
may be introduced by

\begin{defn}[operadic Lax pair \cite{Paal07}]
Allow a classical dynamical system to be described by the
Hamiltonian system \eqref{ham}. An \emph{operadic Lax pair} is a
pair $(L,M)$ of operations $L,M\in\EE_V$, such that the
Hamiltonian system  (\ref{ham}) may be represented as the
\emph{operadic Lax equation}
\[
\frac{dL}{dt}=[M,L]\= M\bul L-(-1)^{|M||L|}L\bul M
\]
The pair $(L,M)$ is also called an \emph{operadic Lax representation} of/for  Hamiltonian system \eqref{ham}.
Evidently, the degree constraints $|M|=|L|=0$ give rise to the ordinary Lax equation (\ref{lax}) \cite{Lax68,BBT03}.
\end{defn}

The Hamiltonian of the harmonic oscillator is
\[
H(q,p)=\frac{1}{2}(p^2+\omega^2q^2)
\]
Thus, the Hamiltonian system of the harmonic oscillator reads
\begin{equation}
\label{eq:h-osc}
\frac{dq}{dt}=\frac{\partial H}{\partial p}=p,\quad
\frac{dp}{dt}=-\frac{\partial H}{\partial q}=-\omega^2q
\end{equation}
If $\mu$ is a linear algebraic operation we can use the above
Hamilton equations to obtain
\[
\dfrac{d\mu}{dt}
=\dfrac{\partial\mu}{\partial q}\dfrac{dq}{dt}+\dfrac{\partial\mu}{\partial p}\dfrac{dp}{dt}
=p\dfrac{\partial\mu}{\partial q}-\omega^2q\dfrac{\partial\mu}{\partial p}
=[M,\mu]
\]
Therefore, we get the following linear partial differential equation for $\mu(q,p)$:
\begin{equation}
\label{eq:diff}
p\dfrac{\partial\mu}{\partial q}-\omega^2q\dfrac{\partial\mu}{\partial p}
=[M,\mu]
\end{equation}
By integrating \eqref{eq:diff} one can get sequences of operations called the
\emph{operadic (Lax representations of) harmonic oscillator}. Since the general solution of the partial differential equations depends on arbitrary functions, these representations are not uniquely determined.

\section{Evolution of binary algebras}

Let $A\= \{V,\mu\}$ be a binary algebra with an operation $xy\=\mu(x\t y)$. 
For simplicity assume that $|M|=0$. 
We require that $\mu=\mu(q,p)$ so that $(\mu,M)$ is an
operadic Lax pair, i.e the Hamiltonian system \eqref{eq:h-osc} of
the harmonic oscillator may be written as  the operadic Lax equation
\[
\dot{\mu}=[M,\mu]\=  M\bul\mu-\mu\bul M,\quad |\mu|=1,\quad |M|=0
\]
Let $x,y\in V$. Assuming that $|M|=0$ and $|\mu|=1$ we have
\begin{align*}
M\bul\mu &=\sum_{i=0}^0M\circ_i\mu
=M\circ_0\mu=M\circ\mu\\
\mu\bul M &=\sum_{i=0}^1\mu\circ_i M =\mu\circ_0 M+\mu\circ_1 M=\mu\circ(M\t\1_V)+\mu\circ(\1_V\t M)
\end{align*}
Therefore,
\[
\dfrac{d}{dt}(xy)=M(xy)-(Mx)y-x(My)
\]
Let $\dim V=n$. In a basis $\{e_1,\ldots,e_n\}$ of $V$,  the
structure constants $\mu_{jk}^i$ of $A$ are defined by
\[
\mu(e_j\t e_k)\=  \mu_{jk}^i e_i,\quad j,k=1,\ldots,n
\]
In particular,
\[
\dfrac{d}{dt}(e_je_k)=M(e_je_k)-(Me_j)e_k-e_j(Me_k)
\]
By denoting $Me_i\=  M_i^se_s$, it follows that
\[
\dot{\mu}_{jk}^i
=\mu_{jk}^sM_s^i-M_j^s\mu_{sk}^i-M_k^s\mu_{js}^i,\quad
i,j,k=1,\ldots, n
\]

\section{Main Theorem}

\begin{lemma}
\label{lemma:harmonic3} Matrices
\[
L\=
\begin{pmatrix}
    p & \omega q & 0 \\
    \omega q & -p & 0 \\
    0 & 0 & 1 \\
  \end{pmatrix}
,\quad
M\=\frac{\omega}{2}
\begin{pmatrix}
    0 & -1 &0\\
1 & 0 & 0\\
0 & 0 & 0
  \end{pmatrix}
\]
give a 3-dimensional Lax representation for the harmonic oscillator.
\end{lemma}

\begin{lemma}
\label{lemma:first}
Let $\dim V=3$ and $M$ be defined as in Lemma
\ref{lemma:harmonic3}. Then the $3$-dimensional binary operadic Lax
equations read
\[
\left\{
  \begin{array}{lll}
\dot{\mu}_{11}^{1}=-\frac{\omega}{2}\left(\mu_{11}^{2}+\mu_{12}^{1}+\mu_{21}^{1}\right),&
\dot{\mu}_{13}^{1}=-\frac{\omega}{2}\left(\mu_{13}^{2}+\mu_{23}^{1}\right),&
\dot{\mu}_{33}^{1}=-\frac{\omega}{2}\mu_{33}^{2}\\
\dot{\mu}_{12}^{1}=-\frac{\omega}{2}\left(\mu_{12}^{2}-\mu_{11}^{1}+\mu_{22}^{1}\right),&
\dot{\mu}_{23}^{1}=-\frac{\omega}{2}\left(\mu_{23}^{2}-\mu_{13}^{1}\right),&
\dot{\mu}_{33}^{2}=\hphantom{-}\frac{\omega}{2}\mu_{33}^{1}\\
\dot{\mu}_{21}^{1}=-\frac{\omega}{2}\left(\mu_{21}^{2}-\mu_{11}^{1}+\mu_{22}^{1}\right),&
\dot{\mu}_{31}^{1}=-\frac{\omega}{2}\left(\mu_{31}^{2}+\mu_{32}^{1}\right),&
\dot{\mu}_{13}^{3}=-\frac{\omega}{2}\mu_{23}^{3}\\
\dot{\mu}_{22}^{1}=-\frac{\omega}{2}\left(\mu_{22}^{2}-\mu_{12}^{1}-\mu_{21}^{1}\right),&
\dot{\mu}_{32}^{1}=-\frac{\omega}{2}\left(\mu_{32}^{2}-\mu_{31}^{1}\right),&
\dot{\mu}_{23}^{3}=\hphantom{-}\frac{\omega}{2}\mu_{13}^{3}\\
\dot{\mu}_{11}^{2}=\hphantom{-}\frac{\omega}{2}\left(\mu_{11}^{1}-\mu_{12}^{2}-\mu_{21}^{2}\right),&
\dot{\mu}_{13}^{2}=-\frac{\omega}{2}\left(\mu_{23}^{2}-\mu_{13}^{1}\right),&
\dot{\mu}_{22}^{3}=\hphantom{-}\frac{\omega}{2}\left(\mu_{12}^{3}+\mu_{21}^{3}\right)\\
\dot{\mu}_{12}^{2}=\hphantom{-}\frac{\omega}{2}\left(\mu_{12}^{1}+\mu_{11}^{2}-\mu_{22}^{2}\right),&
\dot{\mu}_{23}^{2}=\hphantom{-}\frac{\omega}{2}\left(\mu_{23}^{1}+\mu_{13}^{2}\right),&
\dot{\mu}_{21}^{3}=\hphantom{-}\frac{\omega}{2}\left(\mu_{11}^{3}-\mu_{22}^{3}\right)\\
\dot{\mu}_{21}^{2}=\hphantom{-}\frac{\omega}{2}\left(\mu_{21}^{1}+\mu_{11}^{2}-\mu_{22}^{2}\right),&
\dot{\mu}_{31}^{2}=-\frac{\omega}{2}\left(\mu_{32}^{2}-\mu_{31}^{1}\right),&
\dot{\mu}_{11}^{3}=-\frac{\omega}{2}\left(\mu_{21}^{3}+\mu_{12}^{3}\right)\\
\dot{\mu}_{22}^{2}=\hphantom{-}\frac{\omega}{2}\left(\mu_{22}^{1}+\mu_{12}^{2}+\mu_{21}^{2}\right),&
\dot{\mu}_{32}^{2}=\hphantom{-}\frac{\omega}{2}\left(\mu_{32}^{1}+\mu_{31}^{2}\right),&
\dot{\mu}_{12}^{3}=\hphantom{-}\frac{\omega}{2}\left(\mu_{11}^{3}-\mu_{22}^{3}\right)\\
\dot{\mu}_{33}^{3}=0,&
\dot{\mu}_{32}^{3}=\hphantom{-}\frac{\omega}{2}\mu_{31}^{3},&
\dot{\mu}_{31}^{3}=-\frac{\omega}{2}\mu_{32}^{3}
\end{array}
\right.
\]
\end{lemma}
In what follows, consider only anti-commutative algebras. Then one
has
\begin{cor}\label{cor:anticom}
Let $A$ be a 3-dimensional anti-commutative algebra, i.e
\[
\mu^{i}_{jk}=-\mu^{i}_{kj}, \quad i,j,k=1,2,3
\]
Then the operadic Lax equations for the harmonic oscillator read
\[
\left\{
  \begin{array}{lll}
    \dot{\mu}_{12}^{1}=-\frac{\omega}{2}\mu_{12}^{2}, & \dot{\mu}_{12}^{2}=\hphantom{-}\frac{\omega}{2}\mu_{12}^{1}, & \dot{\mu}_{12}^{3}=0\\
    \dot{\mu}_{13}^{1}=-\frac{\omega}{2}\left(\mu_{23}^{1}+\mu_{13}^{2}\right), &
    \dot{\mu}_{13}^{2}=-\frac{\omega}{2}\left(\mu_{23}^{2}-\mu_{13}^{1}\right), &
    \dot{\mu}_{13}^{3}=-\frac{\omega}{2}\mu_{23}^{3}\\
    \dot{\mu}_{23}^{1}=\hphantom{-}\frac{\omega}{2}\left(\mu_{13}^{1}-\mu_{23}^{2}\right), &
    \dot{\mu}_{23}^{2}=\hphantom{-}\frac{\omega}{2}\left(\mu_{13}^{2}+\mu_{23}^{1}\right), &
    \dot{\mu}_{23}^{3}=\hphantom{-}\frac{\omega}{2}\mu_{13}^{3}\\
\end{array}
\right.
\]
\end{cor}
For the harmonic oscillator, define its auxiliary functions $A_\pm$ by
\begin{equation}
\label{eq:def_A}
A_+^2+A_-^2=2\sqrt{2H},\quad
A_+^2-A_-^2=2p,\quad
A_+A_-=\omega q
\end{equation}

\begin{thm}
\label{thm:main} Let $C_{\nu}\in\mathbb{R}$ ($\nu=1,\ldots,9$) be
arbitrary real--valued parameters, such that 
\begin{equation}
\label{eq:cond}
C_2^2+C_3^2+C_5^2+C_6^2+C_7^2+C_8^2\neq0
\end{equation}
Let $M$ be defined as in Lemma
\ref{lemma:harmonic3}, and
\begin{equation}\label{eq:theorem}
\begin{cases}
\mu_{11}^{1}=\mu_{22}^{1}=\mu_{33}^{1}=\mu_{11}^{2}=\mu_{22}^{2}=\mu_{33}^{2}=\mu_{11}^{3}=\mu_{22}^{3}=\mu_{33}^{3}=0\\
\mu_{23}^{1}=-\mu_{32}^{1}=C_2p-C_3\omega q-C_4\\
\mu_{13}^{2}=-\mu_{31}^{2}=C_2p-C_3\omega q+C_4\\
\mu_{31}^{1}=-\mu_{13}^{1}=C_2\omega q+C_3p-C_1\\
\mu_{23}^{2}=-\mu_{32}^{2}=C_2\omega q+C_3p+C_1\\
\mu_{12}^{1}=-\mu_{21}^{1}=C_5A_++C_6A_-\\
\mu_{12}^{2}=-\mu_{21}^{2}=C_5A_--C_6A_+\\
\mu_{13}^{3}=-\mu_{31}^{3}=C_7A_++C_8A_-\\
\mu_{23}^{3}=-\mu_{32}^{3}=C_7A_--C_8A_+\\
\mu_{12}^{3}=-\mu_{21}^{3}=C_9
\end{cases}
\end{equation}
Then $(\mu,M)$ is a $3$-dimensional anti-commutative binary operadic
Lax pair for the harmonic oscillator.
\end{thm}
\begin{proof}
Denote
\[
\begin{cases}
G_{+}^{\omega}\=\;\dot{p}+\omega^{2}q,\quad G_{+}^{\omega/2}\=\dot{A}_++\frac{\omega}{2}A_-\\
G_{-}^{\omega}\=\omega(\dot{q}-p),\quad G_{-}^{\omega/2}\=\dot{A}_--\frac{\omega}{2}A_+\\
\end{cases}
\]
Define the matrix
\[
\Gamma =(\Gamma_{\al}^{\be})\=\begin{pmatrix}
                                   0 & 0 & 0 & 0 & 0 & 0 & 0 & 0 & 0 \\
                                   \hphantom{-}G_{+}^{\omega} & \hphantom{-}G_{+}^{\omega} & G_{-}^{\omega} & G_{-}^{\omega} & 0 & 0 & 0 & 0 & 0 \\
                                   -G_{-}^{\omega} & -G_{-}^{\omega} & G_{+}^{\omega} & G_{+}^{\omega} & 0 & 0 & 0 & 0 & 0 \\
                                   0 & 0 & 0 & 0 & 0 & 0 & 0 & 0 & 0 \\
                                   0 & 0 & 0 & 0 & G_+^{\omega/2} & \hphantom{-}G_-^{\omega/2} & 0 & 0& 0  \\
                                   0 & 0 & 0 & 0 & G_-^{\omega/2} & -G_+^{\omega/2} & 0 & 0 & 0 \\
                                   0 & 0 & 0 & 0 & 0 & 0 & G_+^{\omega/2} & \hphantom{-}G_-^{\omega/2} & 0 \\
                                   0 & 0 & 0 & 0 & 0 & 0 & G_-^{\omega/2} & -G_+^{\omega/2} & 0 \\
                                   0 & 0 & 0 & 0 & 0 & 0 & 0 & 0 & 0 \\
                                 \end{pmatrix}
\]
Then it follows from Corollary \ref{cor:anticom} that the
$3$-dimensional anti-commutative binary operadic Lax equations read
\[
C_{\be}\Gamma_{\al}^{\be}
=C_2\Gamma_{\al}^{2}
+C_3\Gamma_{\al}^{3}
+C_5\Gamma_{\al}^{5}
+C_6\Gamma_{\al}^{6}
+C_7\Gamma_{\al}^{7}
+C_8\Gamma_{\al}^{8}
=0,\quad \al=1,\ldots,9
\]
Since the parameters $C_\be$ ($\be=2,3,5,6,7,8$) are arbitrary, not simultaneously zero, the latter constraints
imply  $\Gamma=0$.

Thus we have to consider the following differential equations
\[
G_{\pm}^{\omega}=0=G_{\pm}^{\omega/2}
\]
We show that
\[
G_{\pm}^{\omega}=0\quad\stackrel{(I)}{\Longleftrightarrow}\quad
\begin{cases}
\dot{p}=-\omega^{2}q\\
\dot{q}=p\\
\end{cases}\quad\stackrel{(II)}{\Longleftrightarrow}\quad
G_{\pm}^{\omega/2}=0
\]

First note that ($I$) immediately follows from the definition of $G_{\pm}^{\omega}$.

The proof of ($II$) can be found in \cite{PV08-1} (Theorem 5.2 ($I$)).
\end{proof}

\section{Initial conditions and dynamical deformations}

It seems attractive to specify the coefficients $C_{\nu}$ in Theorem \ref{thm:main} by the initial
conditions
\[
\left. \mu\right|_{t=0}=\m,\quad
\left. p\right|_{t=0}=p_0\neq0,\quad
\left. q\right|_{t=0}=0
\]
The latter together with \eqref{eq:def_A} yield the initial conditions for $A_{\pm}$:
\[
\begin{cases}
\left.\left(A_+^{2}+A_-^{2}\right)\right|_{t=0}=2\left|p_0\right|\\
\left.\left(A_+^{2}-A_-^{2}\right)\right|_{t=0}=2p_0\\
\left.A_+A_-\right|_{t=0}=0
\end{cases}
\quad \Longleftrightarrow \quad
\begin{cases}
p_0>0\\
\left.A_+\right|_{t=0}=\pm\sqrt{2p_0}\\
\left.A_-\right|_{t=0}=0
\end{cases}
\vee\quad
\begin{cases}
p_0<0\\
\left.A_+\right|_{t=0}=0\\
\left.A_-\right|_{t=0}=\pm\sqrt{-2p_0}
\end{cases}
\]
In what follows assume that $p_0>0$ and $\left.A_+\right|_{t=0}>0$.
Other cases can be treated similarly. 
Note that $p_0=\sqrt{2E}$, where $E>0$ is the total energy of the harmonic oscillator, $H=H|_{t=0}=E$.

From \eqref{eq:theorem} we get the following linear system:
\begin{equation}
\label{eq:constants}
\left\{
  \begin{array}{lll}
    \m{}_{23}^{1}=C_2p_0-C_4, & \m{}_{31}^{1}=C_3p_0-C_1, & \m{}_{12}^{1}=C_5\sqrt{2p_0}\\
    \m{}_{13}^{2}=C_2p_0+C_4, &
    \m{}_{12}^{2}=-C_6\sqrt{2p_0}, &
    \m{}_{23}^{2}=C_3p_0+C_1\\
    \m{}_{13}^{3}=C_7\sqrt{2p_0}, &
\m{}_{23}^{3}=-C_8\sqrt{2p_0}, &
\m{}_{12}^{3}=C_9
\end{array}
\right.
\end{equation}
One can easily check that the latter system can be uniquely solved with respect to $C_\nu$ ($\nu=1,\ldots,9$):
\[
\left\{
  \begin{array}{lll}
C_1=\frac{1}{2}\left(\overset{\circ}{\mu}{}_{23}^{2}-\overset{\circ}{\mu}{}_{31}^{1}\right),&
C_2=\frac{1}{2p_0}\left(\overset{\circ}{\mu}{}_{13}^{2}+\overset{\circ}{\mu}{}_{23}^{1}\right),&
C_3=\frac{1}{2p_0}\left(\overset{\circ}{\mu}{}_{23}^{2}+\overset{\circ}{\mu}{}_{31}^{1}\right)\vspace{1mm}\\
C_4=\frac{1}{2}\left(\overset{\circ}{\mu}{}_{13}^{2}-\overset{\circ}{\mu}{}_{23}^{1}\right),&
C_5=\frac{1}{\sqrt{2p_0}}\overset{\circ}{\mu}{}_{12}^{1},&
C_6=-\frac{1}{\sqrt{2p_0}}\overset{\circ}{\mu}{}_{12}^{2}\vspace{1mm}\\
C_7=\frac{1}{\sqrt{2p_0}}\overset{\circ}{\mu}{}_{13}^{3},&
C_8=-\frac{1}{\sqrt{2p_0}}\overset{\circ}{\mu}{}_{23}^{3},&
C_9=\overset{\circ}{\mu}{}_{12}^{3}
\end{array}
\right.
\]

\begin{rem}
Note that the parameters $C_{\nu}$  have to satisfy condition \eqref{eq:cond} to get the operadic Lax representation for the harmonic oscillaror. 
\end{rem}

\begin{defn}
If $\mu\neq\overset{\circ}{\mu}$, then the multiplication $\mu$ is called a \emph{dynamical deformation} of $\overset{\circ}{\mu}$ (over the harmonic oscillator).
If $\mu=\overset{\circ}{\mu}$, then the multiplication $\overset{\circ}{\mu}$ is called \emph{dynamically rigid}. 
\end{defn}

\section{Examples}

\begin{exam}[$\mathfrak{so}(3)$]
As an example  consider the Lie algebra $\mathfrak{so}(3)$ with the structure equations
\[
[e_1,e_2]=e_3,\quad
[e_2,e_3]=e_1,\quad
[e_3,e_1]=e_2
\]
Thus, the nonzero structure constants are
\[
\m{}_{23}^{1}=\m{}_{31}^{2}=\m{}_{12}^{3}=
-\m{}_{32}^{1}=-\m{}_{13}^{2}=-\m{}_{21}^{3}=1
\]
Using the above initial conditions \eqref{eq:constants}, we get
\[
\left\{
  \begin{array}{lll}
    \m{}_{23}^{1}=C_2p_0-C_4=1, & \m{}_{31}^{1}=C_3p_0-C_1=0, & \m{}_{12}^{1}=C_5\sqrt{2p_0}=0\\
    \m{}_{13}^{2}=C_2p_0+C_4=-1, &
    \m{}_{12}^{2}=-C_6\sqrt{2p_0}=0, &
    \m{}_{23}^{2}=C_3p_0+C_1=0\\
    \m{}_{13}^{3}=C_7\sqrt{2p_0}=0, &
\m{}_{23}^{3}=-C_8\sqrt{2p_0}=0, &
\m{}_{12}^{3}=C_9=1
\end{array}
\right.
\]
From this linear system it is easy to see that the only nontrivial
constants are $C_9=-C_4=1$. Replacing these constants into \eqref{eq:theorem} we get
\[
\mu_{jk}^{i}=\m{}_{jk}^{i},\quad
i,j,k=1,2,3\quad
\Longrightarrow\quad\left.\dot{\mu}\right|_{\mathfrak{so}(3)}=0
\]
Thus we can see that the present selection of the parameters $C_{\nu}$ ($\nu=1,\ldots 9$) via the structure constants of $\mathfrak{so}(3)$ does not give rise to the operadic Lax representation for the harmonic oscillator. Thus $\mathfrak{so}(3)$ is dynamically rigid over the harmonic oscillator. This happens because condition \eqref{eq:cond} is not satisfied.
\end{exam}

\begin{exam}[Heisenberg algebra]
As another example, consider the 3-dimensional Heisenberg algebra $\mathfrak{h}_1$ with the structure equations
%
%
\[
[e_1,e_2]=e_3,\quad [e_1,e_3]=[e_2,e_3]=0
\]
We can see that the only nonzero structure constant is $\m{}_{12}^{3}=1$.
System \eqref{eq:constants} reads
\[
\left\{
  \begin{array}{lll}
    \m{}_{23}^{1}=C_2p_0-C_4=0, & \m{}_{31}^{1}=C_3p_0-C_1=0, & \m{}_{12}^{1}=C_5\sqrt{2p_0}=0\\
    \m{}_{13}^{2}=C_2p_0+C_4=0, &
    \m{}_{12}^{2}=-C_6\sqrt{2p_0}=0, &
    \m{}_{23}^{2}=C_3p_0+C_1=0\\
    \m{}_{13}^{3}=C_7\sqrt{2p_0}=0, &
\m{}_{23}^{3}=-C_8\sqrt{2p_0}=0, &
\m{}_{12}^{3}=C_9=1
\end{array}
\right.
\]
Thus, the only nontrivial constant is $C_9=1$. We conclude that
\[
\mu_{jk}^{i}=\m{}_{jk}^{i},\quad
i,j,k=1,2,3\quad
\Longrightarrow\quad\left.\dot{\mu}\right|_{\mathfrak{h}_1}=0
\]
and ${\mathfrak{h}_1}$ turns out to be dynamically rigid over the harmonic oscillator as well. Again we can see that condition \eqref{eq:cond} is not satisfied.
\end{exam}

\begin{exam}[$\mathfrak{sl}(2)$]
\label{sl2}
Finally consider the Lie algebra $\mathfrak{sl}(2)$ with the structure equations
%
%
\[
[e_1,e_2]=e_3,\quad[e_3,e_1]=2e_1,\quad [e_2,e_3]=2e_2
\]
We can see that the nonzero structure constants are
\[
\m{}_{31}^{1}=\m{}_{23}^{2}=2\m{}_{12}^{3}=2
\]
System \eqref{eq:constants} reads
\[
\left\{
  \begin{array}{lll}
    \m{}_{23}^{1}=C_2p_0-C_4=0, & \m{}_{31}^{1}=C_3p_0-C_1=2, & \m{}_{12}^{1}=C_5\sqrt{2p_0}=0\\
    \m{}_{13}^{2}=C_2p_0+C_4=0, &
    \m{}_{12}^{2}=-C_6\sqrt{2p_0}=0, &
    \m{}_{23}^{2}=C_3p_0+C_1=2\\
    \m{}_{13}^{3}=C_7\sqrt{2p_0}=0, &
\m{}_{23}^{3}=-C_8\sqrt{2p_0}=0, &
\m{}_{12}^{3}=C_9=1
\end{array}
\right.
\]
from which it follows that the only nontrivial constants are $C_3=\frac{2}{p_0},$ $C_9=1$. From 
\eqref{eq:theorem} we get the operadic Lax system
\[
\begin{cases}
\mu_{12}^{1}=\mu_{12}^{2}=\mu_{13}^{3}=\mu_{23}^{3}=\mu_{12}^{3}-1=0\\
\mu_{23}^{1}=\mu_{13}^{2}=-\frac{2\omega}{p_0}q,\quad
\mu_{31}^{1}=\mu_{23}^{2}=\frac{2}{p_0}p
\end{cases}
\]
It turns out that the deformed algebra with multiplication  $\mu$ is also a Lie algebra and isomorphic to $\mathfrak{sl}(2)$. The isomorphism
\[
\mu_{jk}^{s}(q,p)A_s^{i}=\m{}_{lm}^{i}A_j^{l}A_k^{m}
\]
is realized by the matrix
\[
A=(A^{i}_{j}):=
\frac{1}{2p_0}
\begin{pmatrix}
\frac{2p_0}{\omega q}(p+\sqrt{2H}) & 2p_0 & 0 \\
p-\sqrt{2H} & \omega q & 0 \\
0 & 0 & 2\sqrt{2H} \\
\end{pmatrix}
\]
\end{exam}

\section*{Acknowledgement}

The research was in part supported by the Estonian Science Foundation,  Grant ETF 6912.

\medskip
\noindent
Department of Mathematics, Tallinn University of Technology\\
Ehitajate tee 5, 19086 Tallinn, Estonia\\ 
E-mails: eugen.paal@ttu.ee and jvirkepu@staff.ttu.ee

\end{document}